\documentclass[aps,preprint,nobibnotes,nofootinbib]{revtex4}

\usepackage{amsmath}

\begin{document}

\title{Introduction to Gravitational Self-Force}

\author{Robert M. Wald \\ \it Enrico
Fermi Institute and Department of Physics \\ \it University of Chicago
\\ \it 5640 S.~Ellis Avenue, Chicago, IL~60637, USA}

\begin{abstract}
The motion of sufficiently small body in general relativity
  should be accurately described by a geodesic. However, there should
  be ``gravitational self-force'' corrections to geodesic motion,
  analogous to the ``radiation reaction forces'' that occur in
  electrodynamics. It is of considerable importance to be able to
  calculate these self-force corrections in order to be able to
  determine such effects as inspiral motion in the extreme mass ratio
  limit. However, severe difficulties arise if one attempts to
  consider point particles in the context of general relativity. This
  article describes these difficulties and how they have been dealt
  with.
  \end{abstract}

\maketitle

\bigskip
\bigskip

General relativity with suitable forms of matter has a well posed
initial value formulation. In principle, therefore, to determine the
motion of bodies in general relativity---such as binary neutron stars
or black holes---one simply needs to provide appropriate initial data
(satisfying the constraint equations) on a spacelike slice and then
evolve this data via Einstein's equation. It would be highly
desireable to obtain simple analytic descriptions of motion. However,
it is clear that, in general, the motion of a body of finite size will
depend on the details of its composition as well as the details of its
internal states of motion. Therefore, one can hope to get a simple
description of motion only in some kind of ``point particle limit''.
Such a limit encompasses many cases of physical interest, such as
``extreme mass ratio'' inspiral. Of particular interest are the
``radiation reaction'' or ``self-force'' effects occuring during
inspiral---the radiation reaction being, of course, the cause of the inspiral.

By definition, a ``point particle'' is an object whose stress-energy
tensor is given by a delta-function with support on a worldline. A
delta-function makes perfectly good mathematical sense as a
distribution. Now, if a ``source term'' in an equation is
distributional in nature, then the solution to this equation can, at
best, be distributional in nature. Thus, if one wishes to consider
distributional sources, one must generalize the notion of partial
differential equations to apply to distributions.  In the
case of linear equations, this can be done straightforwardly: The
notion of differentiation of distributions is well defined, so it
makes perfectly good mathematical sense to consider distributional
solutions to linear partial differential equations with distributional
sources. Indeed, it is very useful to do so, and, for example, for
Maxwell's equations even if the notion of a ``point charge'' did not
arise from physical considerations, it would be very convenient for
purely mathematical reasons to consider solutions with a
delta-function charge-current source.

However, the situation is different in the case of nonlinear partial
differential equations. Products of distributions normally can only be
defined under special circumstances\footnote{The product of two
  distributions can be defined if the decay properties of their
  Fourier transforms are such that the Fourier convolution integral
  defining their product converges. This will be the case when the
  wavefront sets of the distributions satisfy an appropriate condition
  (see \cite{hormander}).}, so it does not usually even make
mathematical sense to say that a distribution satisfies a nonlinear
equation. Thus, for example, although Maxwell's equations are linear,
the coupled system of Maxwell's equations together with the equations
of motion of the charged matter sources are nonlinear. Consequently,
the complete, ``self-consistent'' Maxwell/motion equations are
nonlinear. Hence, {\it a priori}, these equations are mathematically
ill defined for point charge sources. During the past century, there
has been considerable discussion and debate as to how to make sense of
these equations.

Einstein's equation is nonlinear, so {\it a priori} it does not make
sense to consider this equation with a distributional
source. Nevertheless, it has been understood for the past 40 years
that it does make mathematical sense to consider
Einstein's equation with a shell of matter \cite{israel}, i.e., an object whose
stress-energy tensor is given by a delta-function with support on a
timelike hypersurface (the ``shell''). Solutions to Einstein's
equation with a shell of matter correspond to patching together two
smooth solutions along a timelike hypersurface in such a way that the
metrics induced by the solutions on the two sides of the shell agree,
but the extrinsic curvature is discontinuous. Such a solution
corresponds to having a metric that is continuous, but whose first
derivative has a jump discontinuity across the shell, and whose second
derivative thereby has a delta-function character on the shell. Since
the curvature tensor is linear in the second derivatives of the metric
and there are no terms containing products of first and second
derivatives of the metric, there is no difficulty in making sense of
the curvature tensor of such a metric as a distribution.

Unfortunately, however, the situation is much worse for Einstein's
equation with a point particle source\footnote{``Strings''---i.e.,
objects with a distributional stress-energy tensor corresponding to a
delta-function with support on a timelike surface of co-dimension
two---are a borderline case; see \cite{garfinkle}}. An analysis by 
Geroch and Traschen \cite{gt}
shows that it does not make mathematical sense to consider solutions
of Einstein's equation with a distributional stress-energy tensor
supported on a worldline.  Mathematically, the expected behavior of
the metric near a ``point particle'' is too singular to make sense of
the nonlinear terms in Einstein's equation, even as distributions.
Physically, if one tried
to compress a body to make it into a point particle, it should
collapse to a black hole before a ``point particle limit'' can be
reached.

Since point particles do not make sense in general relativity,
it might appear that no simplifications to the description of motion 
can be achieved. However, the situation is not quite this bad because
it does make mathematical sense to consider 
solutions, $h_{ab}$, to the linearized Einstein equation off of an arbitrary 
background solution, $g_{ab}$, with a
distributional stress-energy tensor supported on a world-line.
Therefore, one might begin a treatment of
gravitational self-force by considering considering solutions to
\begin{equation}
G^{(1)}_{ab}[h](t,x^i) = 8 \pi M u_a(t) u_b(t) \frac{\delta^{(3)}(x^i
  - z^i(t))}{\sqrt{-g}} \frac{d\tau}{dt}\,\, . 
\label{lee}
\end{equation}
Here $u^a$ is the unit tangent (i.e., 4-velocity) of the worldline
$\gamma$ defined by $x^i = z^i(t)$, $\tau$ is the proper time along
$\gamma$, and $\delta^{(3)}(x^i- z^i(t))$ denotes the coordinate
delta-function, i.e., $\int \delta^{(3)}(x^i- z^i(t)) d^3 x^i = 1$.
(The right side of this equation could also be written in a manifestly
covariant form as $8 \pi M \int \delta_4(x,z(\tau))u_a(\tau) u_b(\tau)
d \tau$ where $\delta_4$ denotes the covariant 4-dimensional
delta-function.) However, two major difficulties arise in any
approach that seeks to derive self-force effects starting with the
linearized Einstein equation:
\begin{itemize}
\item 
The linearized Bianchi identity implies that the right side of
eq.(\ref{lee}) must be conserved in the background spacetime. For the
case of a point-particle stress-energy tensor as occurs here,
conservation requires that the worldline $\gamma$ of the particle is a
geodesic of the background spacetime. Therefore, there are no
solutions for non-geodesic source curves, making it a hopeless to use
the linearized Einstein equation to derive corrections to geodesic
motion.
\item
Even if the first problem were solved, solutions to this equation are
singular on the worldine of the particle. Therefore, naive attempts to
compute corrections to the motion due to $h_{ab}$---such as demanding
that the particle move on a geodesic of $g_{ab} + h_{ab}$---are virtually
certain to encounter severe mathematical difficulties, analogous to the
difficulties encountered in treatments of the electromagnetic
self-force problem.
\end{itemize}

The first difficulty has been circumvented by a number of researchers
by modifying the linearized Einstein equation as follows: Choose the
Lorenz gauge condition, so that the linearized Einstein equation
takes the form
\begin{eqnarray}
\label{relaxed}
\nabla^c \nabla_c \tilde{h}_{ab} - 2 R^c{}_{ab}{}^d \tilde{h}_{cd} &=&
- 16 \pi M u_a(t) u_b(t) \frac{\delta^{(3)}(x^i - z^i(t))}{\sqrt{-g}} 
\frac{d\tau}{dt}
\\ \nabla^b \tilde{h}_{ab} &=& 0
\label{lorenz} 
\end{eqnarray}
where $\tilde{h}_{ab} \equiv h_{ab} - \frac{1}{2} h g_{ab}$ with $h =
h_{ab} g^{ab}$.  The first equation, by itself, has solutions for any
source curve $\gamma$; it is only when the Lorenz gauge condition is
adjoined that the equations are equivalent to the linearized Einstein
equation and geodesic motion is enforced. We will refer to the
Lorenz-gauge form of the linearized Einstein equation (\ref{relaxed})
with the Lorenz gauge condition (\ref{lorenz}) {\it not} imposed---as
the {\it relaxed} linearized Einstein equation. If one solves the
relaxed linearized Einstein equation while simply {\it ignoring} the
Lorenz gauge condition that was used to derive this equation, one
allows for the possibility of non-geodesic motion. Of course, the relaxed
linearized Einstein equation is not equivalent to the original
linearized Einstein equation.  However, because deviations from
geodesic motion are expected to be small, the Lorenz gauge violation
should likewise be small, and it thus has been argued that solutions
to the two systems should agree to sufficient accuracy.

In order to overcome the second difficulty, it is essential to
understand the nature of the singular behavior of solutions to the
relaxed linearized Einstein equation on the worldline of the particle.
In order to do this, we would like to have a short distance expansion
for the (retarded) Green's function for a general system of linear wave
equations like (\ref{relaxed}). A formalism for doing this was
developed by Hadamard in the 1920's. It is easiest to explain the
basic idea of the Hadamard expansion in the Riemannian case rather than the
Lorenzian case, i.e., for Laplace equations rather than wave equations. 
For simplicity, we consider a single equation of the form
\begin{equation}
\label{wave}
g^{ab} \nabla_a \nabla_b \phi + A^a \nabla_a \phi + B \phi = 0 \,\, ,
\end{equation}
where $g_{ab}$ is a Riemannian metric, $A^a$ is a smooth vector field,
and $B$ is a smooth function. In the Riemannian case, Green's
functions---i.e., distributional solutions to eq.(\ref{wave}) with a
delta function source on the right side---are unique up to the
addition of a smooth solution\footnote{This follows immediately from
  ``elliptic regularity'', since the difference between two Green's
  functions satisfies the source free Laplace equation (\ref{wave}).},
so all Green's functions have the same singular behavior\footnote{This
  is not true in the Lorentzian case; the singular behavior of e.g.,
  the retarded, advanced, and Feynman propagators are different from
  each other.}.  In $4$-dimensions, in the case where $g_{ab}$ is flat
and both $A^a =0$ and $V=0$, a Green's function with source at $x'$ is
given explicitly by
\begin{equation}
G(x,x') = \frac{1}{\sigma(x,x')}
\end{equation}
where $\sigma(x,x')$ denotes the squared geodesic distance between $x$
and $x'$. This suggests that we seek a solution to the generalized
Laplace equation (\ref{wave}) of the form 
\begin{equation}
G(x,x') = \frac{U(x,x')}{\sigma(x,x')} + V(x,x') \ln \sigma(x,x') + W(x,x')
\label{hform}
\end{equation}
where $V$ and $W$ are, in turn, expanded as
\begin{equation}
\label{series}
V(x,x') = \sum_{j=0}^\infty v_j(x,x') \sigma^j\,\, , \,\,\,\,
W(x,x') = \sum_{j=0}^\infty w_j(x,x') \sigma^j
\end{equation}
One now proceeds by substituting these expansions into the generalized
Laplace equation (\ref{wave}), using the identity $g^{ab} \nabla_a
\sigma \nabla_b \sigma = 4 \sigma$, and then formally setting the
coefficient of each power of $\sigma$ to zero (see, e.g.,
\cite{garabedian}). The leading order equation yields a first order
ordinary differential equation for $U$ that holds along each geodesic
through $x'$. This equation has a unique solution---the square root of
the van Vleck-Morette determinant---that is regular at $x'$. In a
similar manner, setting the coefficient of the
higher powers of $\sigma$ to zero, we get a sequence of ``recursion
relations'' for the quantities $v_j$ and $w_j$, which uniquely determine
them---except for $w_0$, which can be chosen arbitrarily. In the
analytic case, one can then show that the resulting series 
for $V$ and $W$ have a finite radius
of convergence and that the above expansion provides a Green's
function. In the $C^\infty$ but non-analytic case, there is no reason
to expect the series to converge, but truncated or otherwise suitably
modified versions of the series can be used to construct a {\it
  parametrix} for eq.(\ref{wave}). i.e., a solution to eq.(\ref{wave})
with source that differs from a $\delta$-function by 
at most a $C^n$ function.
Even in the analytic case, the Hadamard series defines a
Green's function only in a sufficiently small neighborhood of $x'$.
Clearly, this neighborhood must be contained within a normal 
neighborhood of $x'$ in order that $\sigma$ even be defined.

A similar construction works in the Lorentzian case, i.e., for an
equation of the form (\ref{wave}) with $g_{ab}$ of Lorentz
  signature. The corresponding Hadamard expansion for the retarded
  Green's function is
\begin{equation}
\label{gret}
G_+(x,x') = U(x,x')\delta(\sigma) \Theta(t-t') + V(x,x')
\Theta(-\sigma) \Theta(t-t')
\end{equation}
where $V$ again is given by a series whose coefficients
$v_j$ are uniquely determined by recursion relations. The following
points should be noted

\begin{itemize}
\item
In both the Riemannian and Lorentzian cases, $V(x,x')$ satisfies
eq.(\ref{wave}) in $x$. For a self-adjoint equation (as will be the
case for eq.(\ref{wave}) if $A^a=0$), we also have $V(x,x') = V(x',x)$,
so---where defined---$V$ is a smooth solution of the homogeneous
equation (\ref{wave}) in each variable.

\item 
As already noted above, $v_0(x,x')$ (i.e., the first term in the
series (\ref{series}) for $V$) is uniquely determined by a recursion
relation that can be solved by integrating an ordinary differential
equation along geodesics through $x'$. In particular, in the
Lorentzian case, $v_0(x,x')$ can thereby be obtained on the portion,
$\mathcal{N}$, of the future lightcone of $x'$ lying within a normal
neighborhood of $x'$. Since $\sigma(x,x') = 0$ for $x \in
\mathcal{N}$, we have $V(x,x') = v_0(x,x')$ on $\mathcal{N}$. However,
since---as just noted---$V(x,x')$ satisfies the wave equation in $x$,
it is uniquely determined in the domain of dependence of
$\mathcal{N}$. Thus, in the Lorentzian case, one obtains the form
(\ref{gret}) for the retarded Green's function in a sufficiently small
neighborhood of $x'$ in a way that bypasses any convergence issues 
for the Hadamard series (\ref{series}).

\item
For a globally hyperbolic spacetime, the retarded Green's function
$G_+(x,x')$ is globally well defined. By contrast, as already
emphasized, the Hadamard form (\ref{gret}) of $G_+(x,x')$ can be valid
at best within a normal neighborhood of $x'$. One occasionally sees in
the literature Hadamard formulae that are purported to be valid when
multiple geodesics connect $x$ and $x'$, wherein a summation is made
of contributions of the form (\ref{hform}) or (\ref{gret}) for each
geodesic. I do not believe that there is any mathematical
justification for these formulae.

\item
As follows from the ``propagation of singularities'' theorem \cite{hormander2},
it is rigorously true that, globally, $G_+(x,x')$ is singular if and
only if there is a future directed null geodesic from $x'$ to $x$
(whether or not this geodesic lies within a normal neighborhood of $x'$).

\end{itemize}

Using the generalization of the Hadamard form (\ref{gret}) to the
retarded Green's function for the relaxed linearized Einstein equation
(\ref{relaxed}), one finds (after a very lengthy calculation) that the
solution to eq.(\ref{relaxed}) in a sufficiently small neighborhood of
the point particle source in Fermi normal coordinates is
\begin{equation}
\label{h}
h_{\alpha \beta}=
\frac{2M}{r}\delta_{\alpha \beta} - 8 M a_{(\alpha} u_{\beta)}(1-a_i x^i)
+ h^{\textrm{\tiny tail}}_{\alpha \beta} + M \mathcal{R}_{\alpha \beta} + O(r^2) \,\, .
\end{equation}
Here $r$ denotes the distance to the worldline, $a^\alpha$ is the
acceleration of the worldline, $\mathcal{R}_{\alpha \beta}$ denotes a
term of order $r$ constructed from the curvature of the background
spacetime, and
\begin{equation}
h_{\alpha \beta}^{\textrm{\tiny tail}} \equiv M \int_{-
  \infty}^{\tau^-} \left( G_{+ \alpha \beta \alpha ' \beta '} -
\frac{1}{2} g_{\alpha \beta}G_{+ \ \gamma \alpha ' \beta '}^{\ \gamma
} \right) u^{\alpha '} u^{\beta '}d \tau ' \,\, .
\end{equation}
The symbol $\tau^-$ means that this integration is to be cut short of
$\tau' = \tau$ to avoid the singular behavior of the Green's function
there; this instruction is equivalent to using only the ``tail''
(i.e., interior of the light cone) portion of the Green's function. In
the region where the Hadamard form (\ref{gret}) holds (i.e., for $x$
sufficiently close to $x'$), this corresponds to the contribution to
the Green's function arising from $V(x,x')$.

\bigskip

With the above formula (\ref{h}) for $h_{\alpha \beta}$ as a starting point,
the equations of motion of a point particle---accurate enough to take account
of self-force corrections---have been obtained by the following 3 approaches:

\begin{itemize}
\item
One can proceed in parallel with the derivations of Dirac \cite{dirac}
and DeWitt and Brehme \cite{db} for the electromagnetic case and
derive the motion from conservation of total stress-energy
\cite{mst}. This requires an (ad hoc) regularization of the
``effective stress energy'' associated to $h_{\alpha \beta}$.

\item
One can derive equations of motion from some simple axioms \cite{qw},
specifically that: (i) the difference in ``gravitational force''
between different curves of the same acceleration (in possibly
different spacetimes) is given by the (angle average of the)
difference in $-{\Gamma^\mu}_{\alpha \beta} u^\alpha u^\beta$ where
${\Gamma^\mu}_{\alpha \beta}$ is the Christoffel symbol associated
with $h_{\alpha \beta}$ and (ii) the gravitational self-force vanishes
for a uniformly accelerating worldline in Minkowski spacetime.  This
provides a mathematically clean and simple way of obtaining equations
of motion, but it is not a true ``derivation'' since the motion should
follow from the assumptions of general relativity without having to
make additional postulates.

\item
One can derive equations of motion via matched asymptotic expansions
\cite{mst}, \cite{poisson}.  The idea here is to postulate a suitable
metric form (namely, Schwarzschild plus small perturbations) near the
``particle'', and then ``match'' this ``near zone'' expression to the
``far zone'' formula (\ref{h}) for $h_{\alpha \beta}$. Equations of
motion then arise from the matching after imposition of a gauge
condition. This approach is the closest of the three to a true
derivation, but a number of ad hoc and/or not fully justified
assumptions have been made, most notably Lorentz gauge relaxation.

\end{itemize}

All three approaches have led to the following system of equations (in
the case where there is no ``incoming radiation'', i.e., $h_{ab}$
vanishes in the asymptotic past)
\begin{equation}
\label{misa}
\nabla^c \nabla_c \tilde{h}_{ab} - 2 R^c{}_{ab}{}^d \tilde{h}_{cd} = -
16 \pi M u_a(t) u_b(t) \frac{\delta^{(3)}(x^i - z^i(t))}{\sqrt{-g}}
\frac{d\tau}{dt}
\end{equation}
\begin{equation}
\label{eom}
u^b \nabla_b u^a = - (g^{ab} + u^a u^b)(\nabla_d
h_{bc}^{\tiny \textrm{tail}}- \frac{1}{2} \nabla_b h_{cd}^{\tiny
\textrm{tail}})u^c u^d 
\end{equation}
where it is understood that the retarded solution to the equation for
$\tilde{h}_{ab}$ is to be chosen. Equations (\ref{misa}) and
(\ref{eom}) are known as the MiSaTaQuWa equations.  Note that the
equation of motion (\ref{eom}) for the particle formally corresponds
to the perturbed geodesic equation in the metric $g_{ab} +
h_{ab}^{\tiny \textrm{tail}}$. However, it should be emphasized that
$h_{ab}^{\tiny \textrm{tail}}$ fails to be differentiable on the
worldline of the particle and fails to be a (homogeneous)
solution to the relaxed linearized Einstein equation. (This lack of
differentiablity affects only the spatial derivatives of the spatial
components of $h_{ab}^{\tiny \textrm{tail}}$, so the right side of
eq.(\ref{eom}) is well defined.) Thus, one cannot interpret
$h_{ab}^{\tiny \textrm{tail}}$ as an effective, regularized, perturbed
metric.

An equivalent reformulation of eq.(\ref{eom}) that does admit an
interpretation as perturbed geodesic motion in an effective,
regularized, perturbed metric has been given by Detweiler and Whiting
\cite{dw}, who proceed as follows.  The symmetric Green's function is
defined by $G_{\rm sym} = (G_+ + G_-)/2$ where $G_-$ is the advanced
Green's function. For the case of a self-adjoint wave equation of the
form (\ref{wave})---i.e., with $A^a=0$ and $g_{ab}$ Lorentzian---the
Hadamard expansion of $G_{\rm sym}$ is given by
\begin{equation}
G_{\rm sym}(x,x') = \frac{1}{2} [U(x,x')\delta(\sigma) + V(x,x')
\Theta(-\sigma)] 
\end{equation}
As previously noted $V(x,x')$ is symmetric (i.e., $V(x,x')= V(x',x)$)
and is a homogeneous solution of eq.(\ref{wave}) in each
variable. However, $V(x,x')$ is, at best, defined only when $x$ lies
in a normal neighborhood of $x'$. In the region where $V$ is defined,
Detweiler and Whiting define a new Green's function by
\begin{equation}
G_{\rm DW}(x,x') = \frac{1}{2} [U(x,x')\delta(\sigma) + V(x,x')
\Theta(\sigma)] 
\end{equation}
The Detweiler-Whiting Green's function has the very unusual property
of having no support in the interior of the future or past light
cones. Detweiler and Whiting show that eq.(\ref{eom}) is equivalent to
perturbed geodesic motion in the metric $g_{ab} + h^R_{ab}$ where
$h^R_{ab}$ is obtained by applying $G_+ - G_{\rm DW}$ for the relaxed
linearized Einstein equation to the worldline source. Since $h^R_{ab}$
is a smooth, homogeneous solution to the relaxed linearized Einstein equation,
it can be given an interpretation as an effective, regularized,
perturbed metric. Of course, an observer making spacetime measurements
near the particle would see the metric $g_{ab} + h_{ab}$, not $g_{ab}
+ h^R_{ab}$.

\bigskip

Although there is a general consensus that eqs.(\ref{misa}) and
eq.(\ref{eom}) (or the Detweiler-Whiting version of
eq.(\ref{eom})) should provide a good description of the self-force
corrections to the motion of a sufficiently small body, it is
important that these equations be put on a firmer foundation both to
clarify their range of validity and to potentially enable the
systematic calculation of higher order corrections. It is clear that
in order to obtain a precise and rigorous derivation of gravitational
self-force, it will be necessary to take some kind of ``point particle
limit'',  wherein the size, $R$, of the body goes to zero. However, to
avoid difficulties associated with the non-existence of point
particles in general relativity, it is essential that one let $M$ go
to zero as well. If $M$ goes to zero more slowly than $R$, the body
should collapse to a black hole before the limit $R \rightarrow 0$ is
achieved.  On the other hand, one could consider limits where $M$ goes
to $0$ more rapidly than $R$, but finite size effects would then
dominate over self-force effects as $R \rightarrow 0$.  This suggests
that we consider a one-parameter family of solutions to Einstein's
equation, $g_{ab}(\lambda)$, for which the body scales to zero size
and mass in an asymptotically self-similar way as $\lambda \rightarrow
0$, so that the ratio $R/M$ approaches a constant in the limit. 

Recently, Gralla and I \cite{gw} have considered such one-parameter
families of bodies (or black holes).  In the limit as $\lambda
\rightarrow 0$---where the body shrinks down to a worldline $\gamma$
and ``disappears''---we proved that $\gamma$ must be a
geodesic. Self-force and finite size effects then arise as
perturbative corrections to $\gamma$. To first order in $\lambda$,
these corrections are described by a deviation vector $Z^i$ along
$\gamma$. In \cite{gw}, Gralla and I proved that, in the Lorenz gauge,
this deviation vector satisfies
\begin{equation}
\label{sf}
\frac{d^2Z^i}{dt^2} = \frac{1}{2M} S^{kl} {R_{kl0}}^i - {R_{0j0}}^iZ^j
- \left( {h^{\textrm{\tiny tail}}}^i{}_{0,0} - \frac{1}{2}
h^{\textrm{\tiny tail}}_{00}{}^{,i} \right) 
\end{equation}
The first term in this equation corresponds to the usual ``spin
force'' \cite{papa}, i.e., the leading order finite size correction to
the motion. The second term is the usual right side of the geodesic
deviation equation. (This term must appear since the the corrections
to motion must allow for the possibility of a perturbation to a nearby
geodesic.) The last term corresponds to the self-force term appearing
on right side of eq.(\ref{eom}). It should be emphasized that
eq.(\ref{sf}) arises as the perturbative correction to geodesic motion
for any one-parameter family satisfying our assumptions, and holds for
black holes as well as ordinary bodies.

Although the self-force term in eq.(\ref{sf}) corresponds to the
right side of eq.(\ref{eom}), these equations have different meanings.
Equation (\ref{sf}) is a first order perturbative correction to
geodesic motion, and no Lorenz gauge relaxation is involved in this
equation since $h_{ab}$ is sourced by a geodesic $\gamma$. By
contrast, eq.(\ref{eom}) is supposed to hold even when the cumulative
deviations from geodesic motion are large, and Lorenz gauge relaxation
is thereby essential. Given that eq.(\ref{sf})
holds rigorously as a perturbative result, what is the status of the
MiSaTaQuWa equations (\ref{misa}), (\ref{eom})? In \cite{gw}, we argued
that the MiSaTaQuWa equations arise as ``self-consistent perturbative
equations'' associated with the perturbative result (\ref{sf}). If the
deviations from geodesic motion are {\it locally} small---even though
cumulative effects may yield large deviations from any individual
geodesic over long periods of time---then the MiSaTaQuWa equations
should provide an accurate description of motion.

\bigskip

\noindent {\bf Acknowledgments}

This research was supported in part by NSF grant PHY04-56619 to the
University of Chicago.

\end{document}